\documentclass{aastex62}

\graphicspath{{./}{figures/}}

\begin{document}

\title{Conversion Layer Controls the Evolution of Magnetic Deflections Near the Alfvén Surface}

\correspondingauthor{Dominic Payne}
\email{dspayne@umich.edu}

\author[0000-0002-8650-1310]{Dominic Payne}
\affil{Climate and Space Sciences and Engineering, University of Michigan, Ann Arbor, MI, 48109}

\author{Mojtaba Akhavan-Tafti}
\affiliation{Climate and Space Sciences and Engineering, University of Michigan, Ann Arbor, MI, 48109}

\author{Joshua Goodwill}
\affiliation{Department of Physics and Astronomy, University of Delaware, Newark, DE, 19716}

\author{Samuel Badman}
\affiliation{Center for Astrophysics, Harvard \& Smithsonian, Cambridge, MA, 02138}

\author{Riddhi Bandyopadhyay}
\affiliation{Department of Physics and Astronomy, University of Delaware, Newark, DE, 19716}

\author{Subash Adhikari}
\affiliation{Department of Physics and Astronomy, University of Delaware, Newark, DE, 19716}

\author{William Matthaeus}
\affiliation{Department of Physics and Astronomy, University of Delaware, Newark, DE, 19716}

\author{Gary Zank}
\affiliation{University of Alabama Huntsville, Huntsville, AL, 35899} 

\author{Chen Shi}
\affiliation{Department of Physics, Auburn University, Auburn, AL, 36849}

\author{Michael Stevens}
\affiliation{Center for Astrophysics, Harvard \& Smithsonian, Cambridge, MA, 02138}

\author{Roberto Livi}
\affiliation{University of California Berkeley Space Sciences Laboratory, Berkeley, CA, 94720}

\author{Yeimy Rivera}
\affiliation{Center for Astrophysics, Harvard \& Smithsonian, Cambridge, MA, 02138}

\author{Kristoff Paulson}
\affiliation{Center for Astrophysics, Harvard \& Smithsonian, Cambridge, MA, 02138}

\begin{abstract}
We examine the statistics of Alfvénic deflections in both sub-Alfvénic and super-Alfvénic solar wind with particular focus on a common parameter that underlies the definition of switchbacks: the magnetic deflection angle $\theta_{def}$.  Our findings are in general agreement with earlier studies that suggest magnetic deflection angles of $\theta_{def}>90$ degrees are very unlikely to occur in sub-Alfvénic regimes.   We find that the upper limit of $\theta_{def}$ exhibits an identifiable trend with the Alfvén Mach number $M_a$, suggesting that gradual steepening of Alfvénic deflections with increasing $M_a$ is a plausible mechanism controlling deflection angles in the young solar wind.  Further analysis reveals that large velocity fluctuations ($\delta v / v > 1$) tend to be important in the largest sub-Alfvénic magnetic deflections with increasing contributions from $\delta v_{\parallel}$ very close to $M_a = 1$, while virtually no magnetic deflections in the super-Alfvénic regime exhibit such large velocity perturbations.  We also determine the local ratio of radial Poynting flux $S_R$ to kinetic energy flux $K_R$ and find that large sub-Alfvénic deflection angles tend to be dominated by $S_R$, while super-Alfvénic deflections are eventually dominated by the $K_R$ associated with the radial solar wind flow.  Our results show that within the vicinity of the Alfvén surface (where $M_a=1$), there is a critical region of parameter space within which $\delta v \sim v_a$ and $K_R/S_R \sim 1$. We refer to this region (where $|\log_{10}(M_a)| < 0.2$) as the conversion layer.  The conversion layer may play a significant role in the evolution of magnetic defections by providing the medium for converting magnetic energy to particle energy and likely driving the formation of magnetic switchbacks in super-Alfvenic solar wind.

\end{abstract}

\section{Introduction} \label{sec:intro}
Ever since the Parker Solar Probe (PSP) observed localized Alfvénic structures with correlated radial velocity spikes and magnetic field reversals in the young solar wind \citep{bale2019highly,kasper2019alfvenic}, the origin location and generation mechanism of these so-called ``switchbacks" have been topics of interest to the heliophysics community.  One potentially relevant threshold responsible for switchback generation is the Alfvén critical surface $R_a$ \citep{chhiber2024alfven}, where the ratio of local solar wind bulk velocity to the local Alfvén speed (also known as the Alfvénic mach number $M_a$) is equal to one.  Results from recent encounters of the sub-Alfvénic solar wind ($M_a <1$) suggest that switchbacks cease to exist below $R_a$ \citep{bandyopadhyay2022sub,akhavan2024situ}, and therefore switchbacks may be the result of an in-situ generation mechanism that manifests at or beyond the $M_a=1$ threshold \citep{akhavan2024situ}.   

It is not yet clear whether full switchbacks form at the Alfvén surface or whether they form gradually from other types of Alfvénic deflections that do not meet more constrained definitions of switchbacks until they have propagated beyond $R_a$.  If magnetic deflections exceeding a 90 degree deflection angle can form gradually from initially sub-Alfvénic magnetic deflections below the 90 degree threshold, then the switchback generation mechanism could occur well below $R_a$, forming a seed structure that ultimately steepens into a mature switchback beyond $R_a$.  

In the following study, we use a variety of intervals from encounters 13-23 to examine the characteristics of magnetic deflections, particularly as a function of $M_a$ to explore what role the $M_a = 1$ threshold plays in switchback formation in the solar wind. This study is complementary to and was conducted in collaboration with \cite{goodwill_swtichback_2025}, who examined the statistical dependence between $M_a$, the switchback parameter (related to the magnetic deflection angle) and deflections in radial velocity and magnetic field.  Although this study and that of \cite{goodwill_swtichback_2025} differ in the range of PSP encounters used and the specific methods for identifying deflections, both studies approach similar conclusions regarding the overall trends with respect to $M_a$.  Section \ref{sec:method} describes the instruments used for data collection, definitions of key parameters, and the criteria used to select and filter deflection data.  In section \ref{sec:angs}, we present our initial findings of the distribution of magnetic deflection angles as a function of $M_a$ and radial distance from the sun.  Section \ref{sec:vels} examines the velocity deflections associated with the distribution of magnetic deflections as well as the ratio of radial energy flux densities, and how these parameters differ in sub- vs super-Alfvénic regimes and with magnetic deflection angle.  In section \ref{sec:discussion}, we discuss the implications of our results for switchback formation and finally summarize our findings in section \ref{sec:conclusion}.

\section{Data and Methods} \label{sec:method}
The magnetic field data in this study come from the FIELDS \citep{bale2016fields} instrument suite on PSP, which provides vector magnetic field measurements up to a 290 sample per second cadence.  The ion velocity data comes from the SWEAP instrument suite on PSP \citep{kasper2016solar}, specifically the Solar Probe Analyzer for ions \citep{livi2022solar}.  The density is determined via the quasi-thermal noise (QTN), which comes from the Radio Frequency Spectrometer (RFS) included with FIELDS \citep{pulupa2017solar}.  All data products used in this study are interpolated to match the time cadence of SPAN-I velocity data before any variables or time-averages are computed.  

Our objective is to explore how Alfvénic deflections systematically differ for a range of $M_a$ values that span the $M_a=1$ threshold.  $M_a$ is calculated based on the mean proton velocity magnitude $\langle|v_p|\rangle$ and the Alfvén speed $v_a = \frac{\langle|B|\rangle}{\sqrt{\mu_0 m_i n}}$ where $\langle|B|\rangle$ is the mean magnetic field magnitude, $\mu_0$ is the vacuum magnetic permeability, $m_i$ is the proton mass,  $n$ is the density derived from QTN, and angled brackets indicate a 10 minute average.  
$$  M_a = \frac{\langle|v_p|\rangle}{v_a}$$

From a variety of PSP perihelia we select intervals that are either sub-Alfvénic ($M_a<1$) or super-Alfvénic ($M_a>1$) for at least 6 consecutive hours and we include an additional collection of ``near-Alfvénic" intervals where $M_a$ is close to unity to get sufficient data close to the Alfvén surface (see supplemental material).  

Some studies tend to have more strict definitions of switchbacks \citep{akhavan2021discontinuity}, while others tend to only consider the deflection angle as the defining feature \citep{de2020switchbacks}.  We do not exclude deflection angles below 90 degrees from our selections.  However, we do filter out any uncorrelated magnetic and velocity fluctuations with the condition that both $\delta v/v>0.05$  and $\delta B/B>0.05$.  The $\delta B/B$ term is a scalar quantity defined explicitly below.
$$\frac{\delta B}{B} =\frac{|\vec{B} - \vec{\langle B \rangle}|}{|\vec{\langle B \rangle}|}$$
where $\vec{B}$ is the vector magnetic field (interpolated to the SPAN-I velocity data cadence) and $\vec{\langle B \rangle}$ is the mean vector magnetic field using a 10 minute averaging window.  $\delta v/v$ is defined in the same way, using the vectors $\vec{v}$ and $\vec{\langle v \rangle}$.  The magnetic deflection angle $\theta_{def}$ is the angle between the vector magnetic field measurement and the mean vector magnetic field over the 10 minute averaging window, defined below.
$$\theta_{def} = \arccos\Bigg(\frac{  \vec{B}\cdot \vec{\langle B \rangle}}{|\vec{B}| \    |\vec{\langle B \rangle}|  }\Bigg)$$

The following work also examines the statistics of the energy flux densities associated with the distribution of magnetic deflections with varied $M_a$.  The two most relevant to the the solar wind are the Poynting flux $\vec{S}$ associated with the transport of electromagnetic energy and the ion kinetic energy flux $\vec{K}$ associated with the bulk flow of ion motion.  The radial components of each are expressed below.  
$$S_R = \frac{(\vec{E}\times\vec{B})_R }{\mu_0}$$
$$\vec{K}_R = \frac{n_i m_i v^2_R}{2}\vec{v}_R$$
To approximate the electric field in the Poynting flux calculation, we determine the convective electric field using the ion bulk velocity.
$$\vec{E} = \vec{E}_{convective} = -(\vec{v}_i \times\vec{B})$$
In the following sections, we compare a sub- and super-Alfvénic interval, investigate the distribution of magnetic deflection angles with $M_a$, and examine their velocity and energy flux characteristics.

\section{Comparing a Sub-Alfvénic and Super-Alfvénic Interval}

\begin{figure}
    \centering
    \includegraphics[scale=0.5]{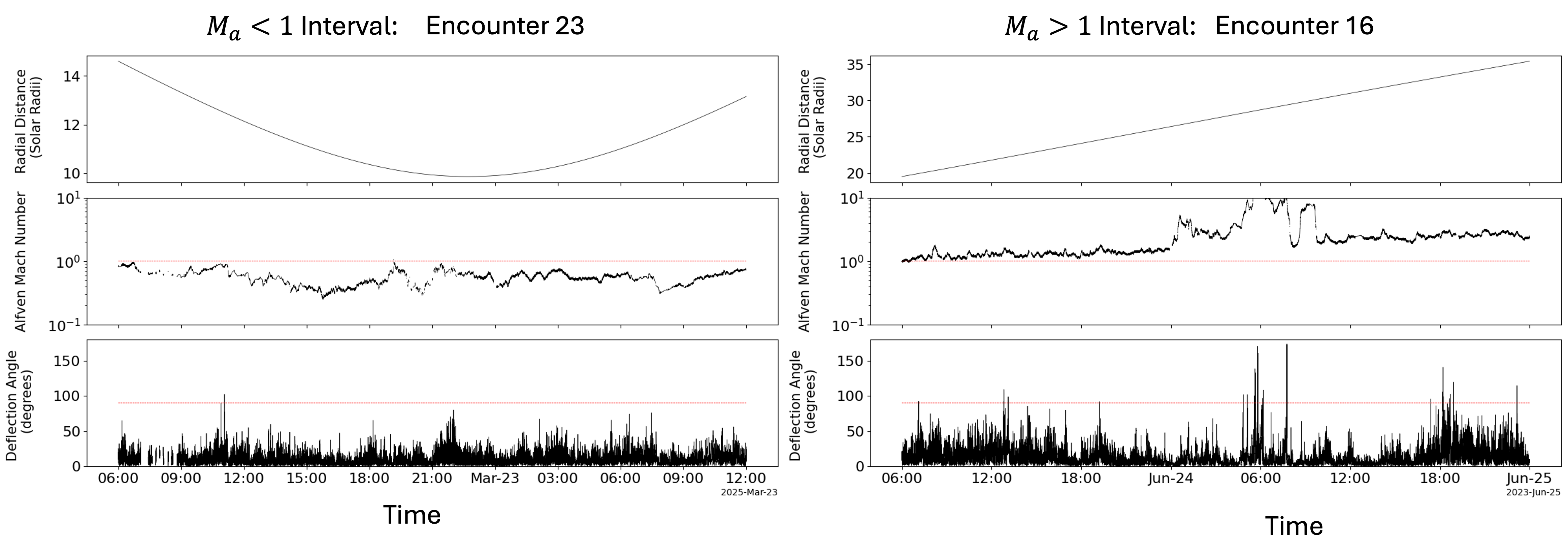}
    \caption{Interval Examples: (a) Sub-Alfvénic Interval from encounter 23 and (b) super-Alfvénic interval from encounter 16.  Each example shows, from top to bottom, the heliocentric distance of PSP in units of solar radii, the Alfvénic Mach number with a red dotted line at $M_a = 1$, and the deflection angle with a red dotted line at $\theta_{def} = 90$ degrees.}
    \label{fig:timeseries}
\end{figure}

Figure \ref{fig:timeseries} shows examples of a sub-Alfvénic and a super-Alfvénic PSP interval from encounters 23 and 16, respectively. In each set of panels we show the distance of PSP from the sun, the Alfvénic Mach number, and the magnetic deflection angle ($\theta_{def}$).  The super-Alfvénic interval has multiple periods with $\theta_{def}>90$ degrees, while in this particular sub-Alfvénic interval $\theta_{def}$ rarely exceeds 90 degrees.  While this could imply a relationship between $M_a$ and $\theta_{def}$, it is necessary to investigate the relationship between these two variables across many different encounters spanning a range of $M_a$ values.

\section{Distribution of Deflection Angles} \label{sec:angs}

\begin{figure}
    \centering
    \includegraphics[scale=0.5]{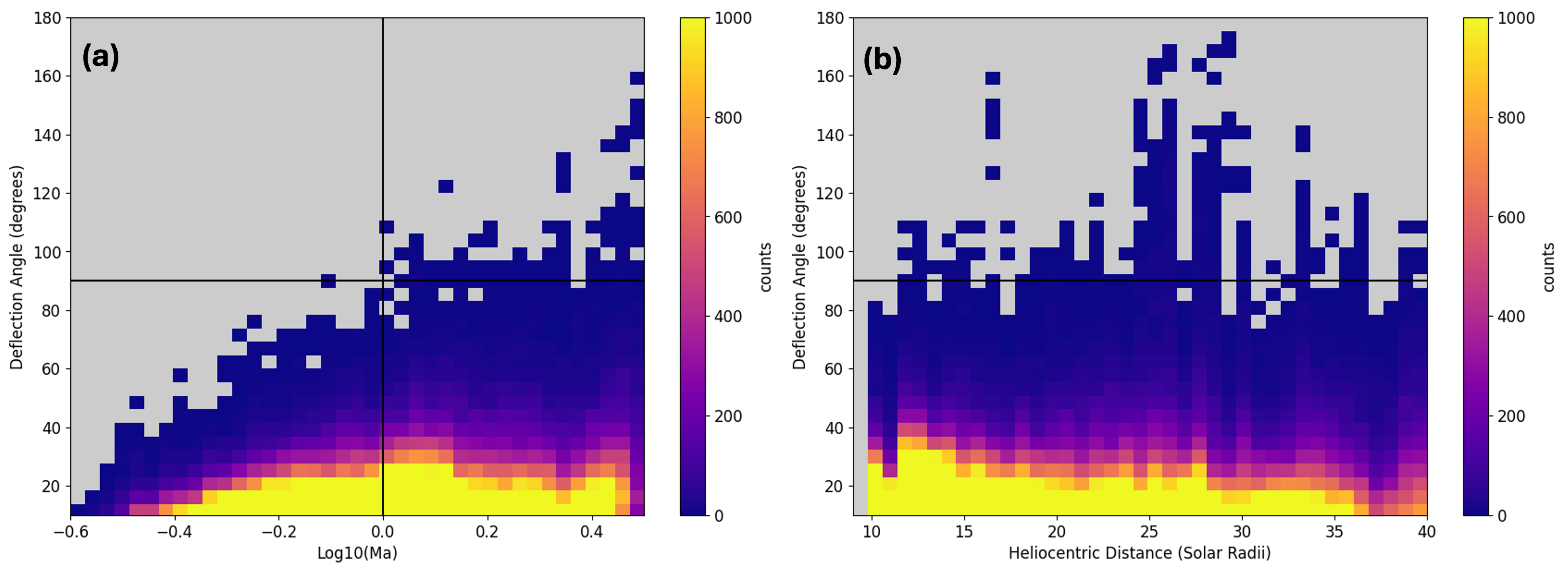}
    \caption{Distribution of Magnetic deflection angles (a) as a function of $\log_{10}(M_a)$ with black lines indicating $M_a =1$ and $\theta_{def} = 90$ degrees and (b) as a function of Heliocentric distance with a horizontal black line indicating $\theta_{def} = 90$ degrees. }
    \label{fig:angles}
\end{figure}

The distribution of $\theta_{def}$ as a function of $\log_{10}(M_a)$ is presented as a 2D histogram in figure \ref{fig:angles}.  The lack of significant  $\theta_{def} >90 $ degree data in the sub-Alfvénic regime ($\log_{10}(M_a) < 0$) is consistent with earlier studies \cite{bandyopadhyay2022sub,akhavan2024situ,adhikari2025characterization} that relied on the earliest sub-Alfvénic encounters by PSP, suggesting that full switchbacks in the sub-Alfvénic regime are exceedingly rare compared to super-Alfvénic intervals.  The distribution of the largest $\theta_{def}$ with increasing $M_a$ appears fairly continuous, suggesting that magnetic deflection angles tend to become systematically larger with increasing $M_a$, no matter their originating mechanism.  When plotting as a function of distance from the sun, rather than $M_a$, there is not as obvious a trend, though there is a notable increase in larger deflections beyond $\sim 25$ solar radii.  

\section{Properties of Sub- and Super-Alfvénic Velocity Deflections} \label{sec:vels}

\begin{figure}
    \centering
    \includegraphics[scale=0.5]{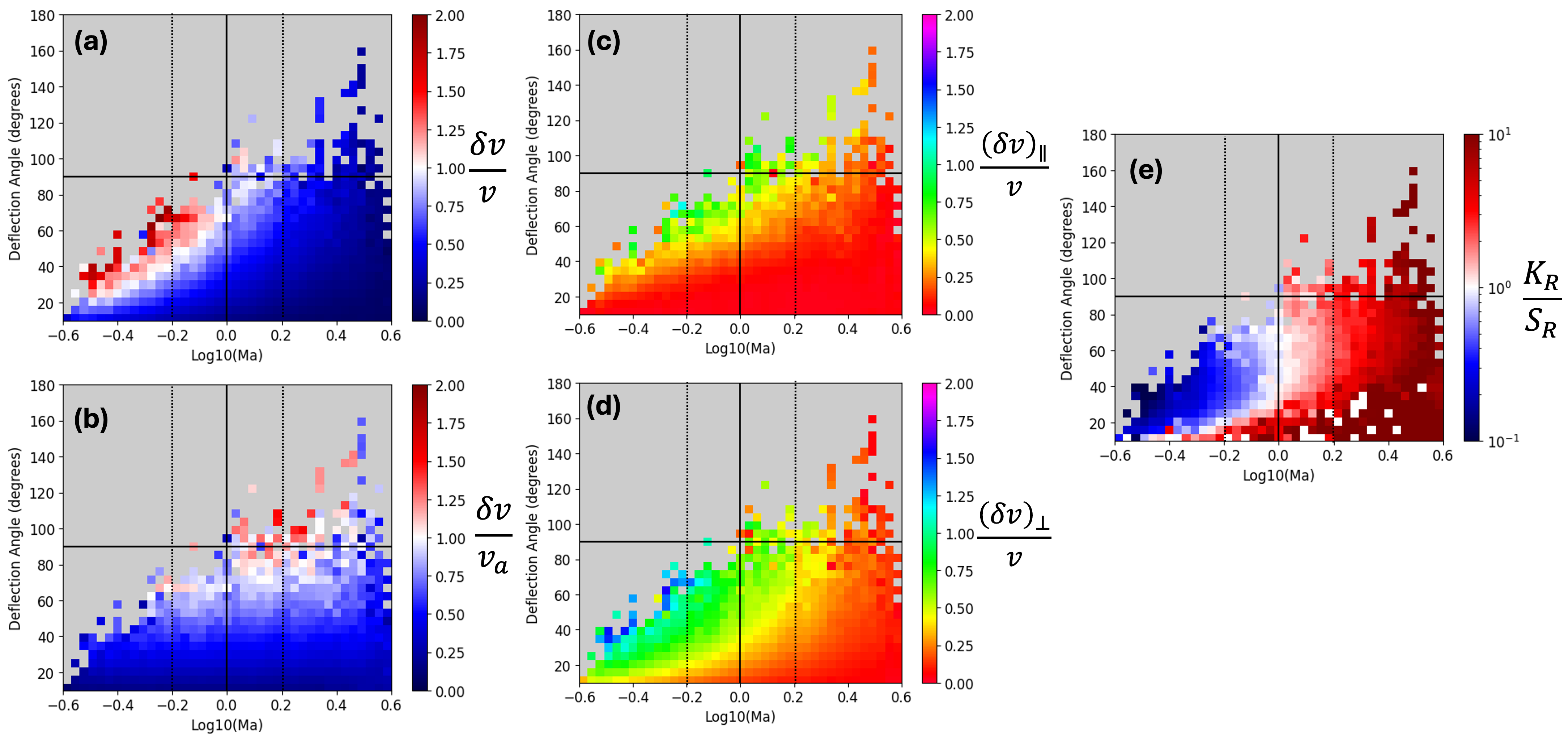}
    \caption{Distribution of velocity deflections with respect to $\log_{10}(M_a)$ and deflection angle normalized to (a) the mean bulk velocity and (b) the mean Alfvén speed. Also included are separate velocity deflection contributions (c) parallel and (d) perpendicular to the mean magnetic field, each normalized to the mean bulk velocity.  (e) Ratio of radial ion kinetic energy flux to radial Poynting flux with respect to $\log_{10}(M_a)$ and deflection angle.  The solid black lines in all plots indicate the values of $\log_{10}(M_a) = 0$ (or $M_a = 1$) and $\theta_{def} = 90$ degrees.  The vertical dotted lines in all plots indicate values of $|\log_{10}(M_a)| = 0.2$ as a visual aid.}
    \label{fig:dv}
\end{figure}

To further examine the nature of the sub- to super-Alfvénic transition, we explore the properties of velocity fluctuations and energy flux densities associated with the deflections in each regime. Figure \ref{fig:dv}(a-d) again shows 2D histograms of the distribution of deflection angles vs $M_a$, with $\delta v/v$, $\delta v/v_a$, $(\delta v)_\parallel /v$, and $(\delta v)_\perp /v$ represented by the color scales.  In the sub-Alfvénic regime, the largest $\theta_{def}$ values tend to be associated with strong velocity deflections that exceed the bulk velocity magnitude ($\delta v/v > 1$), while the largest deflections in the super-Alfvénic regime tend to have $\delta v/v \lesssim 1$.  When normalized to the Alfvén speed, $\delta v / v_a$ rarely exceeds unity in the sub-Alfvénic regime, but $\delta v / v_a \sim 1$ close to the Alfvén surface in the sub-Alfvénic regime, followed by more $\delta v / v_a > 1$ data at large deflection angles in the super-Alfvénic regime.  Figure \ref{fig:dv}(c-d) uses a different color scale to better emphasize how the different components of $\delta v$ vary in parameter space.  The $(\delta v)_\perp /v$ contributions tend to dominate and are at their most significant for the largest values of $\theta_{def}$ deep in the sub-Alfvénic regime where $\log_{10}(M_a) \lesssim -0.2$.  In contrast, the $(\delta v)_\parallel /v$ term tends to be at its largest when $|\log_{10}(M_a)|<0.2$.

The distribution of $\delta v/v$ and its component contributions suggest that within the sub-Alfvénic regime, there is some variability in the nature of the Alfvénic fluctuations as $M_a$ approaches unity.  In figure \ref{fig:dv}e we examine the energy flux content associated with these fluctuations based on the distribution of $K_R/S_R$. The larger $\theta_{def}$ values in the sub-Alfvénic regime tend to be dominated by $S_R$ while the smaller ones (below $\sim20$ degrees)  tend to have larger $K_R$. From $\log_{10}(M_a) \simeq -0.2 - 0$,  the the $K_R/S_R$ ratio approaches unity for the largest $\theta_{def}$.  In the super-Alfvénic regime, $K_R$ tends to dominate at nearly all deflection angles, especially for $\log_{10}(M_a)>0.2$.

\section{Discussion} \label{sec:discussion}

We have analyzed the statistics of magnetic deflection angles, velocity deflections, and energy flux densities from a variety of intervals spanning radial distances of $\sim 9.8 - 40$ solar radii with significant data coverage in the range $M_a \sim 0.4 - 2.5$, thus including the $M_a=1$ threshold that defines the Alfvén surface.  Having a variety of radial distances and $M_a$ values is important since the physical location of the Alfvén surface is highly variable \citep{cranmer2023sun, chhiber2022extended}, and $M_a$ should not be treated as a direct proxy for radial distance from the solar surface, even if $M_a$ increases with radial distance generally. 

The distribution of the largest magnetic deflections (figure \ref{fig:angles}) systematically increases with $M_a$ and that the largest deflections reach the threshold of $\theta_{def} > 90$ degrees for $\log_{10}(M_a)\ge0$.  These results imply that the upper limit of $\theta_{def}$ is somewhat controlled by the local $M_a$ within and near the corona.  They also imply that magnetic switchbacks, often defined in part by $\theta_{def}>90$ degrees \citep{akhavan2024situ}, could  form through a steepening process that affects Alfvénic structures above and below it.  This is consistent with the theory \citep{toth2023theory} that switchback formation is a result of the distortion of spherically polarized Alfvén waves by transverse gradients in the wave speed, which can gradually increase $\theta_{def}$ up to and beyond 90 degrees during propagation.  

The velocity fluctuations exhibit distinct characteristics across the sub- to super-Alfvénic transition.  It is clear from figure \ref{fig:dv}a that the largest magnetic deflection angles in the sub-Alfvénic regime tend to have velocity fluctuations exceeding the mean bulk velocity ($\delta v/v >1$), while there are very few cases of $\delta v/v>1$ in the super-Alfvénic regime.  Figure \ref{fig:dv}b shows that $\delta v / v_a$ approaches unity in the range $\log_{10}(M_a) \sim -0.2 - 0$, but rarely exceeds unity until $\log_{10}(M_a) >  0$. These striking results suggest that near the Alfvén surface, the $\delta v/v_a \sim 1$ criterion may create instabilities that tend to erode large velocity fluctuations ($\delta v/v >1$) in the super-Alfvénic regime.  This is consistent with \cite{ruffolo2020shear}, who showed that velocity shears near the Alfvén speed allow Kelvin-Helmholtz instabilities to develop when magnetic tension becomes too weak to suppress them, ultimately enhancing turbulence and magnetic deflection angles beyond the Alfvén surface.  

Decomposition of the separate components of $\delta v/v$ in figure \ref{fig:dv}(c-d) show that the influence of $(\delta v)_\perp$ tends to dominate, especially for the largest deflection angles deep in the sub-Alfvénic regime ($\log_{10}(M_a)<-0.2$).  For the largest magnetic deflections near the Alfvén surface ($|\log_{10}(M_a)|<0.2$), the normalized $(\delta v)_\parallel$ fluctuations are at their largest and roughly comparable to the $(\delta v)_\perp$ components.  The significant influence of $(\delta v)_\parallel$ in this critical $|\log_{10}(M_a)|<0.2$ range suggests a departure from purely Alfvénic characteristics to include more magnetosonic behavior consistent with earlier studies of switchback characteristics \citep{zank2020origin}. 

The radial energy flux ratio $K_R/S_R$ also exhibits dependencies with $M_a$ and $\theta_{def}$.  It is clear from figure \ref{fig:dv}e that in the sub-Alfvénic regime, most of the magnetic deflections have $S_R > K_R$ and $K_R/S_R$ is smaller for larger $\theta_{def}$.  The reduced influence of $S_R$ at small $\theta_{def}$ makes sense considering that in the $\theta_{def} \sim 0$ limit (assuming background magnetic field dominated by $B_R$), there are no $B_T$ or $B_N$ components that can contribute to $S_R$.  However, in the range $\log_{10}(M_a) \sim -0.2-0$ there is also some reduced influence at the largest values of $\theta_{def}$ where the $K_R/S_R \lesssim 1$.  From $\log_{10}(M_a) \sim 0-0.2$, the contributions from $K_R$ and $S_R$ are roughly comparable for $\theta_{def} \sim 40-80$ degrees, but dominated by $K_R$ for $\theta_{def}\lesssim40$ degrees and $\theta_{def}\gtrsim80$ degrees.  $K_R$ dominates for all deflection angles beyond $\log_{10}(M_a) \sim 0.2$.  A simple analysis can help make sense of the result of this transition with $M_a$.  If we take $K \sim nmv^3/2$ and $S \sim EB/\mu_0 \sim vB^2 / \mu_0$, then $K/S \sim \mu_0nmv^2/2B^2 = v^2/2v_a^2 = M_a^2/2$.  Setting $K/S = 1$ and solving for $M_a$ we obtain $M_a = \sqrt{2}$ or $\log_{10}(M_a) \simeq 0.15$, which is close to $\log_{10}(M_a)\sim 0.2$ beyond which $K_R$ starts to dominate for all deflections.  

\begin{figure}
    \centering
    \includegraphics[width=0.6\linewidth]{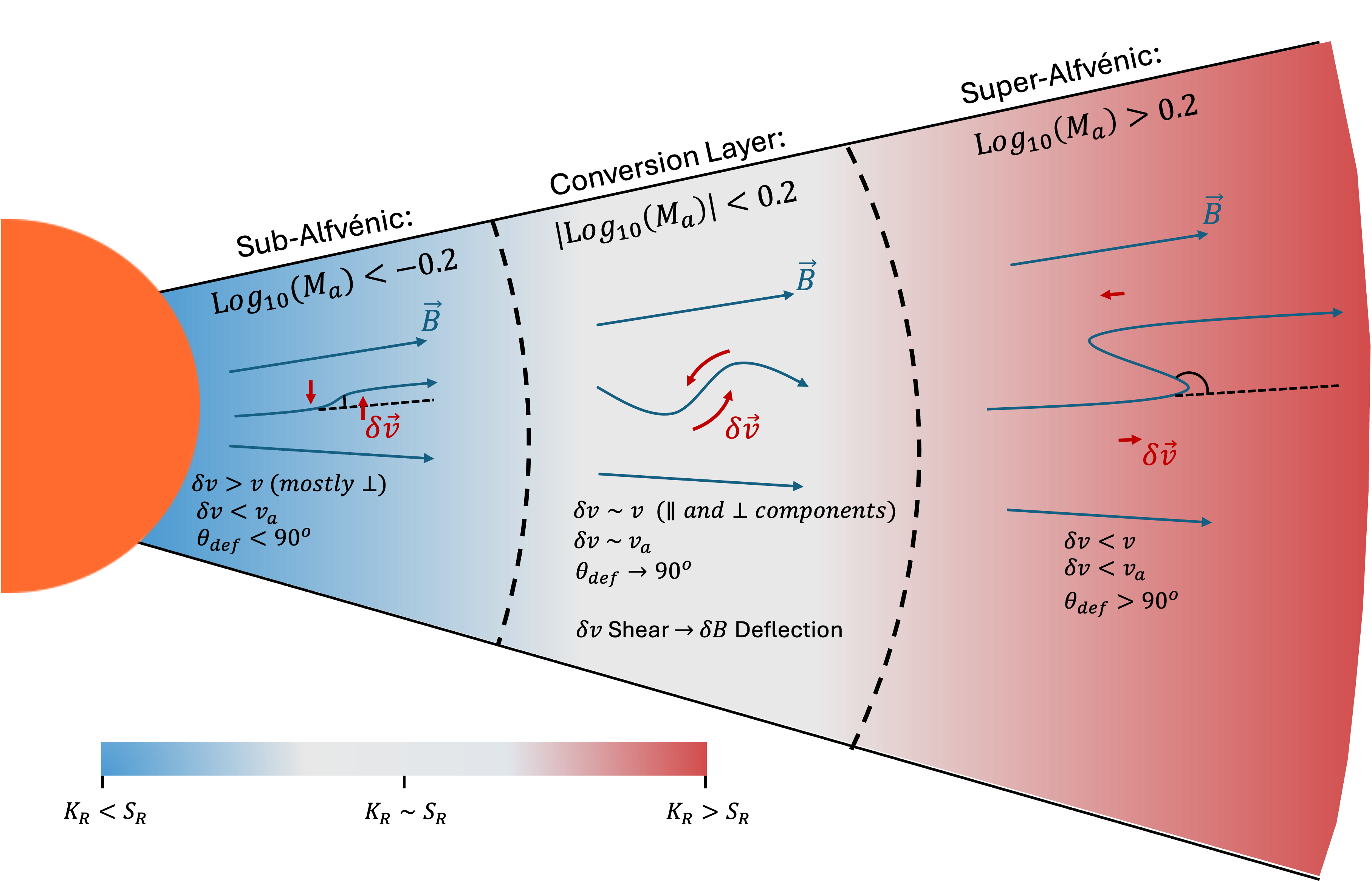}
    \caption{Diagram illustrating the characteristics of the sub-Alfvénic and super-Alfvénic regimes as well as the so-called 'conversion layer' where proximity to the Alfvén surface introduces nonlinear interactions between magnetic and velocity deflections.  }
    \label{fig:diagram}
\end{figure}

These statistical results do not necessarily imply that an individual structure moving across a $M_a=1$ boundary will suddenly lose most of its velocity shear and electromagnetic energy flux content.  However they do imply that Alfvénic deflections tend to undergo a transition in their velocity energy flux characteristics within a conversion layer where $|\log_{10}(M_a)|\lesssim 0.2$. We present a diagram in figure \ref{fig:diagram} that illustrates the main results of the preceding sections and the observed characteristics of the sub-Alfvénic regime, the super-Alfvénic regime, and the conversion layer (not to scale). Large velocity deflections in the conversion layer become increasingly parallel to the magnetic field and become unstable as $\delta v$ approaches the Alfvén speed.  As this occurs, instabilities like those described in \cite{ruffolo2020shear} can arise which result in the termination of large velocity deflections and can produce large magnetic deflection angles during KH-type roll up processes.  Meanwhile, the radial components of the Poynting flux and ion kinetic energy flux densities become comparable in the conversion layer before kinetic energy flux fully dominates in the super-Alfvénic regime. 

In conclusion, our study shows that the conversion layer encompassing the Alfvén surface likely plays a role in the formation of magnetic switchbacks \citep{akhavan2024situ} and contributes to the acceleration and heating of solar wind \citep{akhavan2022magnetic, mostafavi2025preferential}.

Before closing, the authors note a potential challenge to the stated findings of this study; namely the $\theta_{def}>90$ degree deflection near 11:00 UT in figure \ref{fig:timeseries}, which appears to occur in a sub-Alfvénic interval.  Upon close inspection of that interval (not shown) we find that there is a small gap in the QTN density data, and therefore the derived $M_a$, concurrent with this deflection, which explains why this particular sub-Alfvénic deflection exceeding 90 degrees does not show up in the statistical results.  Alternate methods for determining $M_a$ to fill in the gap such as smoothing or using SPAN-I density do seem to indicate that the interval could be a genuine outlier where $\theta_{def}<90$ degrees and $M_a < 1$, but verification and further analysis of this event is beyond the scope of this study.  We emphasize that the statistical results of this study do not rule out the possibility of $\theta_{def}<90$ degree deflections in the sub-Alfvénic regime, but they do suggest such cases are exceedingly rare compared to the super-Alfvénic regime.

\section{Conclusion}\label{sec:conclusion}
We have examined the statistics of magnetic deflection angles in a variety of sub-Alfvénic and super-Alfvénic solar wind regimes.  Our initial finding is that while the distribution of magnetic deflection angles as a function of heliocentric distance may be complicated, the maximum observed magnetic deflection angles are somewhat tied to the local Alfvén Mach number. This demonstrates that large magnetic deflections can result from a steepening process that gradually increases small magnetic deflection angles during propagation eventually resulting in full reversals in super-Alfvénic solar wind.  The velocity fluctuations associated with the largest sub-Alfvénic deflection angles tend to exceed the background flow magnitude, but such large velocity deflections tend not to survive into the super-Alfvénic regime, even for the largest magnetic deflection angles.  Further decomposition of the parallel and perpendicular components of the velocity deflections in the sub-Alfvénic regime suggests that while they are mainly dominated by perturbations perpendicular to the local magnetic field, parallel perturbations play an increasing role as the sub-Alfvénic mach number approaches unity, suggesting a mixture of Alfvénic and magnetosonic characteristics.  The distribution of the energy flux ratio suggests that below the Alfvén surface, the Poynting flux density tends to dominate for all but the smallest deflection angles, but the kinetic energy flux dominates for $\log_{10}(M_a)>0.2 $.  These results demonstrate that while full switchback production may not be entirely localized precisely at a $M_a = 1$ boundary, there exists what we refer to as the conversion layer ($|\log_{10} (M_a)| \lesssim 0.2$) which contributes to switchback formation and field-particle energy conversion.

\section*{Acknowledgments}
This work was primarily supported by NASA contract number NNN06AA01C.  This work would not be possible without the entirety of the PSP team, including those that keep the instrumentation operational and those involved in data management.
\section*{Data Availability}
The PSP data used in this study is publicly available from the Coordinated Data Analysis Web (https://cdaweb.gsfc.nasa.gov/pub/data/psp/).

\bibliography{Alfven}

\vspace{5mm}





\end{document}